# Foamed emulsion drainage: flow and trapping of drops

Maxime Schneider*, Ziqiang Zou, Dominique Langevin and Anniina Salonen*

Laboratoire de Physique des Solides, CNRS, Univ. Paris-Sud,
Universite Paris-Saclay, 91405 Orsay Cedex, France.
E-mail: maxime.schneider@u-psud.fr, anniina.salonen@u-psud.fr

**Abstract :** Foamed emulsions are ubiquitous in our daily life but the ageing of such systems is still poorly understood. In this study we investigate foam drainage and measure the evolution of the gas, liquid and oil volume fractions inside the foam. We evidence three regimes of ageing. During an initial period of fast drainage, both bubbles and drops are very mobile. As the foam stabilises drainage proceeds leading to a gradual decrease of the liquid fraction and slowing down of drainage. Clusters of oil drops are less sheared, their dynamic viscosity increases and drainage slows down even further, until the drops become blocked. At this point the oil fraction starts to increase in the continuous phase. The foam ageing leads to an increase of the capillary pressure until the oil acts as an antifoaming agent and the foam collapses.

## 1 Introduction

Foamed emulsions or foamulsions are made of gas bubbles and oil drops dispersed in water.[1,2] Bubbles in an aqueous surfactant foam are shown in *Fig. 1(a)* and in a foamed emulsion in *Fig. 1(b)*. The emulsion drops are found in the spaces between the bubbles and can change the properties and feel of the resulting foam. These foams are encountered in many food products, such as whipped creams, mousses or ice creams,[3] where some (or all of) the oil is crystallized at the bubble surfaces to provide stability[4]. On the market there are whipped cosmetics products also, such as creams, foundations and cleansing products, where some of the components are oil-based. Oil foams being notoriously difficult to stabilize [5,6], emulsification is an interesting alternative that offers the possibility to obtain stable aqueous foams containing significant amounts of oil. Foamulsions have already been shown to be useful precursors for the generation of porous materials, where the non-aqueous phase is solidified through polymerization [7,8] or carbonization [9].

The formulation of foamed emulsions is not always simple as many oils are anti-foams and destroy foams rapidly [10] However, when there exists a barrier to the entry of the oil drops into the gas–water interfaces, the foamed emulsions can be very stable [1,2,11]. Koczo et al.[1] studied the ageing of foamed emulsions with different oils, oil volume fractions and oil drop sizes. They showed that the emulsion drops can increase the life-time of the foams considerably by slowing down drainage. The foam stabilization depends on the drop size via different mechanisms: larger drops are blocked in Plateau borders and concentrate inside the foam, while smaller drops act by increasing the viscosity of the continuous phase. Intermediate drop sizes give the least stable foams. Highly concentrated emulsions can have a yield stress that is sufficient to stop drainage [11,12]. Goyon et al.[11] showed that drainage could be induced by shear once the applied stress exceeds the yield stress of the emulsions. Similarly we showed that drainage will restart if the foam coarsens, the bubble size increases and with it the dimensions of the Plateau borders and the gravitational stress [2]. There exists a number of studies, where foam drainage is studied in the presence of solid particles. In the case of solid particles, Pitois and co-workers have also shown that how the drainage proceeds depends strongly on the ratio of the particle size to the foam channel size [13–16].

The stability of foamed emulsions can be modulated by changing the type of stabilizer [17–19], as in oil-free foams, and by the choice of oil. Foamulsions with longer alkanes have been shown to be more stable [1,19] than those made from more soluble oils. The dual nature of the drops as stabilisers and antifoamobjects has even been used to make stimulable foamed emulsions, where they have been destroyed using UV-light or temperature [18,20].

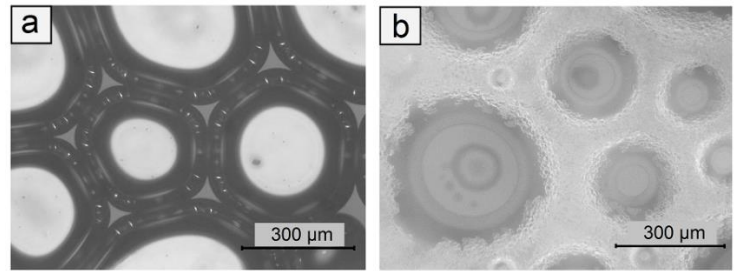

*Fig. 1 (a)* Picture of a pure SDS foam taken with an optical microscope right after generation. The liquid fraction is about 0.15. *(b)* Picture of a foamed emulsion made from an emulsion with a 0.3 oil fraction in the continuous phase and at the same foam liquid fraction.

In this paper we address the question of foamed emulsion drainage. The three phases of the foamed emulsion (water with surfactant, oil and gas) will separate over time, we quantify the evolution of the three fluid fractions inside the foams. Experiments at different heights inside the foam and with different emulsions show that drainage proceeds in steps, which depend on the properties of the emulsion.

## 2 Materials and methods

### 2.1 Materials: surfactant, oil and gas

For all experiments, a single surfactant is used, sodium dodecylsulfate (SDS, fromSigma-Aldrich) that stabilizes both the emulsion drops and the foam bubbles. The SDS solutions were used at most one day after preparation. Emulsions are oil in water dispersions made of rapeseed oil (Rapeseed oil from Brassica Rapa, from Sigma-Aldrich) and MilliQ water (conductivity 18.2 M$\Omega$ cm). Rapeseed oil is a mixture of triglycerides, also containing lesser amounts of mono- and di-glycerides.

The oil has been chosen because it is not a strong antifoaming agent. This can be evaluated considering the entry (E), bridging (B) and spreading (S) coefficients [10,21], which are reported in *Table 1*. The entry coefficient E is positive if the oil dispersed in water is able to penetrate into the air–water interface. The bridging coefficient B is positive if the oil drops are able to bridge the foam films. Finally, the spreading coefficient S is positive if the oil can spread at the air–water interface. Antifoam action is optimal if all these coefficients are positive. They can be evaluated from the interfacial tensions between oil and water $\gamma_{OW}$, air and water $\gamma_{AW}$ and air and oil $\gamma_{AO}$, which are also shown in *Table 1*.

The interfacial tensions were measured using a drop (bubble) tensiometer (Tracker, Teclis, France) in pendant (for $\gamma_{AO}$) or rising drop (for $\gamma_{AW}$ and $\gamma_{OW}$) operating modes. Images were then analyzed using the drop shape analysis software. One sees in Table 1 that B is positive meaning that oil drops can bridge the two sides of foam films. However, E and S, although positive, are small so that rapeseed oil is not a strong antifoaming agent with the surfactant used. In addition, we use very high concentrations of SDS (30 g L$^{-1}$), so the droplet and bubble surfaces are well covered by surfactant and the entry barrier is rather high: this barrier prevents oil from entering the air–water interface even if E is positive [10].

| $\gamma_{AW}$ | $\gamma_{OW}$ | $\gamma_{AO}$ | E | B | S |
|---|---|---|---|---|---|
| 36 | 2 | 33 | 5 | 211 | 1 |

*Table 1* Interfacial tensions between oil and water $\gamma_{OW}$, air and water $\gamma_{AW}$ and air and oil $\gamma_{AO}$, entry (E), bridging (B) and spreading (S) coefficients. Units are mN m$^{-1}$ except for B, given in (mN m$^{-1})^2$

## 2.2 Emulsion preparation

Emulsions are prepared with an oil volume fraction of 70% by mixing rapeseed oil and an SDS solution at 30 g L$^{-1}$ using two different methods: the so-called double-syringe technique [22] and sonication. The resulting emulsions are referred to as SG or SN respectively. For both techniques, oil and surfactant solutions are pre-mixed using a spatula. The double-syringe technique makes use of two 50 mL syringes from CODAN Medical and a plastic tube junction with an inner diameter of 4.3 mm. The liquids are circulated between syringes 20 times to generate the emulsion. The sonicator (Ultrasonic processor, Bioblock scientific with a 19 mm bar) was operated during 5 min at 20 kHz, 130 W, amplitude of 90% and alternating 1 s of pulse with 2 s of rest to avoid heating up the solution. The emulsions are diluted with the SDS solution to obtain different oil volume fractions f ranging from 10 to 40%. The size and polydispersity of the emulsion drops is measured using a Malvern Mastersizer laser granulometer (Malvern Instruments, France) which is based on Mie scattering, using a refractive index for oil noil = 1.47. Emulsions have been diluted in a SDS solution for proper background subtraction. The emulsions' polydispersity is around 40% with the two generation methods.

## 2.3 Foam generation

The emulsions prepared as described in *Section 2.2* were foamed using the double syringe technique, allowing us to control easily the initial liquid fraction of the foam, without changing the bubble size distribution [22]. To measure the bubble size distribution, a sample of foam is collected and then diluted in a 10 g L$^{-1}$ SDS solution (enough to prevent bubble coalescence). The resulting mean bubble radius is determined by optical microscopy, averaging over at least 1 000 bubbles.

## 2.4 Measuring creaming of the emulsion

To estimate the creaming velocity of the emulsions, we have put the different emulsions in small tubes (height of 12 cm) and we have taken pictures of the rising front between clear fluid and concentrated emulsion which moves upward at a constant velocity in the initial stages of creaming.

## 2.5 Monitoring foamulsion ageing

For all experiments, the column is filled using two 50 mL syringes full of foam that are connected through a T-junction. Both syringes are emptied simultaneously into the column. In order to follow the detail of the drainage in foamed emulsions, the evolution of the bubble and drop concentrations have to be measured quantitatively. We measured the water fraction $\varepsilon_W$ and the liquid fraction of the foam $\varepsilon$ from which we can deduce the fraction of oil in the foam $\varepsilon_O$. If $\varepsilon_A$ is the air fraction, the different liquid fractions of the foam are linked in the following way:

$$\varepsilon + \varepsilon_A = 1$$
$$\varepsilon = \varepsilon_O + \varepsilon_W \tag{1}$$

A commercially available device Foamscan (Teclis, France) is used to measure the normalized conductivity of the foamed emulsion $\Lambda = \frac{\sigma_{foam}}{\sigma_W}$, linked to $\varepsilon_W$ through a semi-empirical relation [23] :

$$\varepsilon_W = \frac{3\Lambda(1 + 11\Lambda)}{1 + 25\Lambda + \Lambda^2} \tag{2}$$

The FoamScan by Teclis has been customized to acquire images of the surface of the column at the same position as the conductivity probes, as shown in the drawing of *Fig. 2(a)*. Typical images are shown in *Fig. 2(b and c)*. Image analysis allows us to determine the surface liquid fraction $\varepsilon_s$. The liquid fraction of the foam $\varepsilon$ can be calculated from $\varepsilon_s$ by [24]:

$$\varepsilon = 0.64\left(\sqrt{1-\varepsilon_s} - 1\right)^2 \tag{3}$$

Threshold determination to binarize the images is a crucial step in image processing. Since bubbles are very small at the beginning and since the average brightness of the images increases while the foam ages, it is important to treat images individually as seen in the images in *Fig. 2(b and c)*. In order to validate the procedure, we checked that the initial liquid fraction is the same as the one chosen during foam generation.

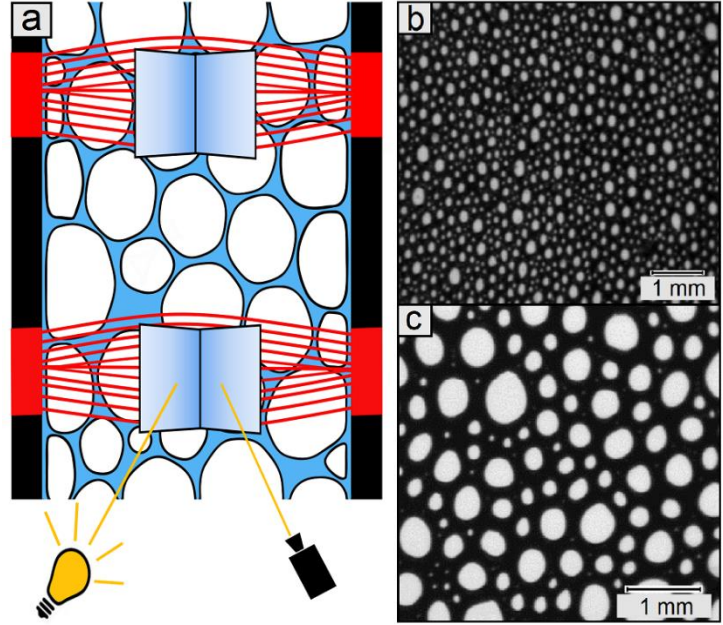

***Fig. 2 (a)*** *Experimental setup: foam conductivity is measured at different heights z while pictures of the surface are taken at these heights through a prism with a telecentric lens mounted on a camera.* ***(b)*** *Picture of the surface for a foamed emulsion made from an emulsion with 10% of oil right after generation.* ***(c)*** *Picture of the surface of the same foam after 8 min.*

Evolution of average bubble radii over time has been determined from the surface pictures. The "Analyze Particles" function from Fiji (an open source image processing package based on ImageJ) is used to determine the number of bubbles in an image. The total surface is divided by this number to obtain the average surface per bubble. This number is then converted into an average bubble radius ($R = \sqrt{\frac{A}{\pi}}$, R being the average bubble radius and A being the average bubble surface). This is an approximation but gives an idea of the evolution of the bubble radius over time. The difference between the liquid fraction of the foam e and the conducting fraction of the foam $\varepsilon_W$, is the oil fraction in the foam $\varepsilon_O$ (rapeseed oil conductivity is much smaller than that of water). We define the oil fraction f in the liquid as:

$$\phi = \frac{\varepsilon_O}{\varepsilon} = \frac{\varepsilon - \varepsilon_W}{\varepsilon} \tag{4}$$

Calibrations have been made to ensure the applicability of *Eqn (3)* on both pure SDS foam and foamed emulsions. Further tests were made by measuring independently the concentration of oil inside the foam by extracting samples of the foam at a given height, as done by Koczo et al. in Ref. [1]. The results are shown in the ESI† in *Fig. S2* and the agreement between measurements is very good.

# 3 Theoretical background

## 3.1 Creaming and drainage

Air bubbles and oil drops rise in water because of density differences. For an isolated drop, the rising velocity is the Stokes velocity $v_{Stokes}$ given in *Eqn (5)*:

$$v_{Stokes} = \frac{2\Delta\rho g r_H^2}{9\eta} \quad (5)$$

where $\Delta\rho$ is the density difference between water and oil, $r_H$ is the drop hydrodynamic radius, $\eta$ the liquid viscosity, g the gravitational acceleration. The emulsions have drop volume fractions in the range of 10–40% and the creaming velocity $v_c$ is less than that given by *Eqn (5)* due to hydrodynamic interactions between drops. An empirical formula known to describe well sedimentation of concentrated particles writes:

$$v_c = v_{Stokes}(1 - \phi)^n \quad (6)$$

The exponent n depends on the Reynolds number [25]; for low Reynolds numbers, n is usually between 4.65 and 6.55. *Eqn (6)* has already been used to account for the creaming velocity of emulsions [26].

## 3.2 Drainage of aqueous foams

In foams, the bubbles are packed together and the rising velocity is much smaller than predicted by the Stokes formula. It is more common to talk about drainage of liquid rather than rising of bubbles. The liquid is contained in the films between bubbles, Plateau borders or nodes. The amount of liquid in the films is usually negligible and the problem can be viewed as the flow of liquid within the network of Plateau borders [27]. Drainage proceeds in two steps. In the first step, the liquid fraction starts to decrease from the top of the foam. The limit between this drier foam zone and the wetter foam zone below is the drainage front, which moves downwards at a constant velocity V:

$$V = \frac{K_d \rho g L^2 \varepsilon^\alpha}{\eta} \quad (7)$$

where $K_d$ is a foam permeability, L the length of Plateau borders (PBs) and a an exponent depending on the surface mobility: $\alpha = 1$ for rigid surfaces (liquid friction dominated by the PBs), $\alpha = 0.5$ for mobile surfaces (liquid friction dominated by the nodes). In this paper, we used sodium dodecyl sulfate (SDS) as foam stabilizing agent which forms air–water interfaces of high mobility, hence $\alpha \sim 0.5$ [28]. In the second step, the foam continues to drain more slowly until the equilibrium vertical liquid fraction profile is attained when the hydrostatic pressure is equilibrated by the osmotic pressure in the foam [29]. Once this is reached, drainage stops (assuming bubble size no longer changes).

## 3.3 Drainage of particle-laden foams

The drainage of particle-laden foams can be quite different from simple aqueous foams if the size or concentration of particles is sufficiently large. In particular the ratio of the size of the particles to the size of the Plateau borders l is critical to determining how drainage proceeds. This ratio was named confinement parameter by O. Pitois and colleagues [14–16,30], and is a key parameter to know whether the particle suspension or the pure liquid flows. l is defined as:

$$\lambda = \frac{r_P}{R_{PB}} \quad (8)$$

where $r_p$ is the particle radius and $R_{PB}$ the average Plateau border radius. For λ < 1, the particles are free to drain (flowing suspension) and for λ > 1, the particles are trapped and their mobility is strongly reduced (flowing liquid).

We will use the same expression to estimate the Plateau border radius as Louvet et al. [30] given by:

$$R_{PB} = \frac{0.27\sqrt{\varepsilon} + 3.17\varepsilon^{2.75}}{1 + 0.57\varepsilon^{0.27}} R \quad (9)$$

This result follows from a Surface Evolver calculation and is valid for liquid fractions below 0.26, as in all our experiments.

### 3.4 Flocculation in emulsions

When the aqueous continuous phase of emulsions contains surfactant micelles, these micelles give rise to an attractive interaction between oil drops due to the depletion of micelles in the spaces between drops. If the attraction is sufficiently strong, flocculation can occur. When the micellar radius rmic is much smaller than oil drop radius r, as in the work here, the depletion energy can be written as [31] :

$$U_{dep} = -1.5 k_B T \phi_m \frac{r}{r_{mic}} \quad (10)$$

where $k_B$ is the Boltzmann constant, $\phi_m$ is the micelle volume fraction and T the absolute temperature. Here $r_{mic} \sim$ 2.5 nm and r ~ 1 μm; *Eqn (10)* shows that for even small volume fractions of micelles, the energy $U_{dep}$ can be well above $k_B T$.

## 4 Results

### 4.1 Characterization of emulsions and foamed emulsions

The average oil drop radius in SN emulsions is about 0.3 μm and in SG emulsions about 1.2 μm. The emulsions are stable for months, during which the radius and polydispersity remain constant. The viscosities of the different emulsions are presented in ESI† (see *Fig. S1*). The emulsions are shear thinning, but freely flowing fluids (shear modulus G0 = 0).

The initial bubble radii are presented in Table 2. The bubble size is around 35 μm initially except for SN30 and SG40 which are more viscous and hence make smaller bubbles (around 20 μm in radius [22]). The polydispersity is set by the generation method and is around 40% for all foamed emulsions.

|         | Φ = 0.1 | Φ = 0.2 | Φ = 0.3 | Φ = 0.4 |
|---------|---------|---------|---------|---------|
| SN      |         |         |         |         |
| R [μm]  | 36.2    | 38.6    | 32.5    | 26.7    |
| Poly.   | 0.38    | 0.38    | 0.41    | 0.44    |
| SG      |         |         |         |         |
| R [μm]  | 35.5    | 39.5    | 38.1    | 26.0    |
| Poly.   | 0.36    | 0.34    | 0.37    | 0.34    |

*Table 2* Bubble radius and polydispersity immediately after generation for the different foams studied

## 4.2 Creaming of the emulsion

We have measured the creaming velocity of the different emulsions used in this study by following the rise of the creaming front in tubes from photographs such as shown in the inset of *Fig. 3*. The initial variation of the creaming front position is linear in time and allows us to calculate the creaming velocities, which are shown as a function of oil volume fraction f in *Fig. 3*. SG emulsions have a clear creaming front between water and concentrated emulsion whereas SN emulsions' creaming front is less clear as the bottom part is light gray, suggesting that some small drops are left behind.

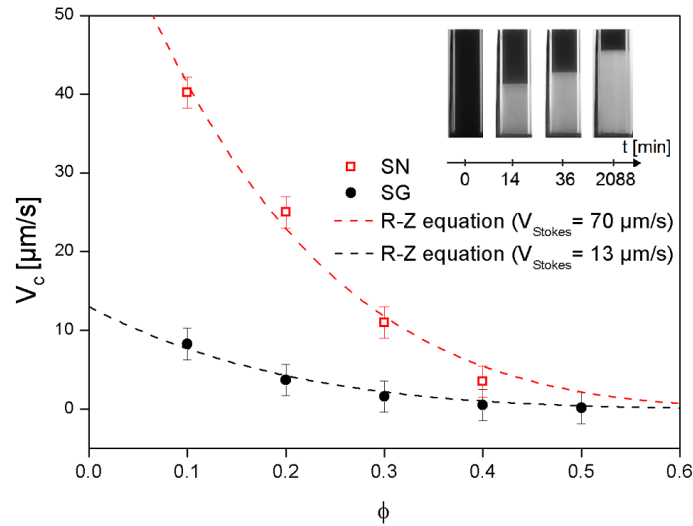

*Fig. 3* Creaming velocity of SG (emulsion made using the double syringe technique) SN (emulsion made by sonication) emulsions respectively red open squares and black circles as a function of oil volume fraction and fits with Eqn (6). SN emulsions have a more pronounced decrease in velocity as we increase the oil fraction than SG emulsions. Top right corner: Pictures taken in transmission of the creaming emulsions at different times: due to density difference, oil is rising towards the top of the tubes.

For both emulsions the creaming velocity decreases with the oil volume fraction, as expected from *Eqn (6)*. However, the creaming velocities are much higher than expected for the size of droplets that we have. Estimation of the creaming velocity can be done using *Eqn (5)*, V ~ 0.28 µm $s^{-1}$ for SG and V ~ 0.02 µm $s^{-1}$ for SN emulsions. This suggests that the drops are flocculated because of the high concentration of SDS. In order to estimate the size of the flocculated entities, we fitted the creaming data using *Eqn (6)*, taking n = 5 (dashed lines in *Fig. 3*) and obtained an effective Stokes velocity for each emulsion. Guignot et al. estimated the size of colloidal clusters sedimenting inside a foam in a similar manner [32]. The fits are rather good indicating that a single effective hydrodynamic radius is a reasonable assumption. *Eqn (5)* was then used to obtain the effective hydrodynamic radius, $r_H$. The drops do not coalesce, so $r_H$ can be considered as a cluster radius. We need the density of the clusters to estimate their size, and we assume random close packing in the clusters (volume fraction of 64%). The cluster size for SG emulsions is about 12 µm while for SN emulsions it is about 28 µm, so much larger than the average individual drop radii which are 1.2 and 0.3 µm respectively. The obtained cluster radii are estimations and are mainly used to be able to calculate the confinement parameter l. We can estimate the interaction energy using *Eqn (10)*, as 12 $k_BT$ for SG and 3 $k_BT$ for SN taking 30 g $L^{-1}$ SDS, i.e. $\phi_{mic}$ = 3%. The values are high and explain the clustering of the drops.

## 4.3 Foam drainage: different emulsion volume fractions

We have followed the evolution of the different fluid fractions and the change in the bubble size over time at a fixed position in the column (z/H = 0.66) for an initial liquid fraction $\varepsilon_i$ = 0.15. The oil volume fraction in the emulsion continuous phase was varied from 0.1 to 0.4 and two different emulsion drop sizes were used (SG of 1.2 μm and SN of 0.3 μm). We show in *Fig. 4* the evolution of fluid fractions and of bubble size over time.

The accuracy of liquid fraction measurements is within a few percent due to uncertainties in liquid volume measurements in the two-syringe method and due to threshold limitations during image processing. Hence the measured $\varepsilon_i$ can differ from the set value of 0.15 (up to 0.18 as seen in *Fig. 4(a)*).

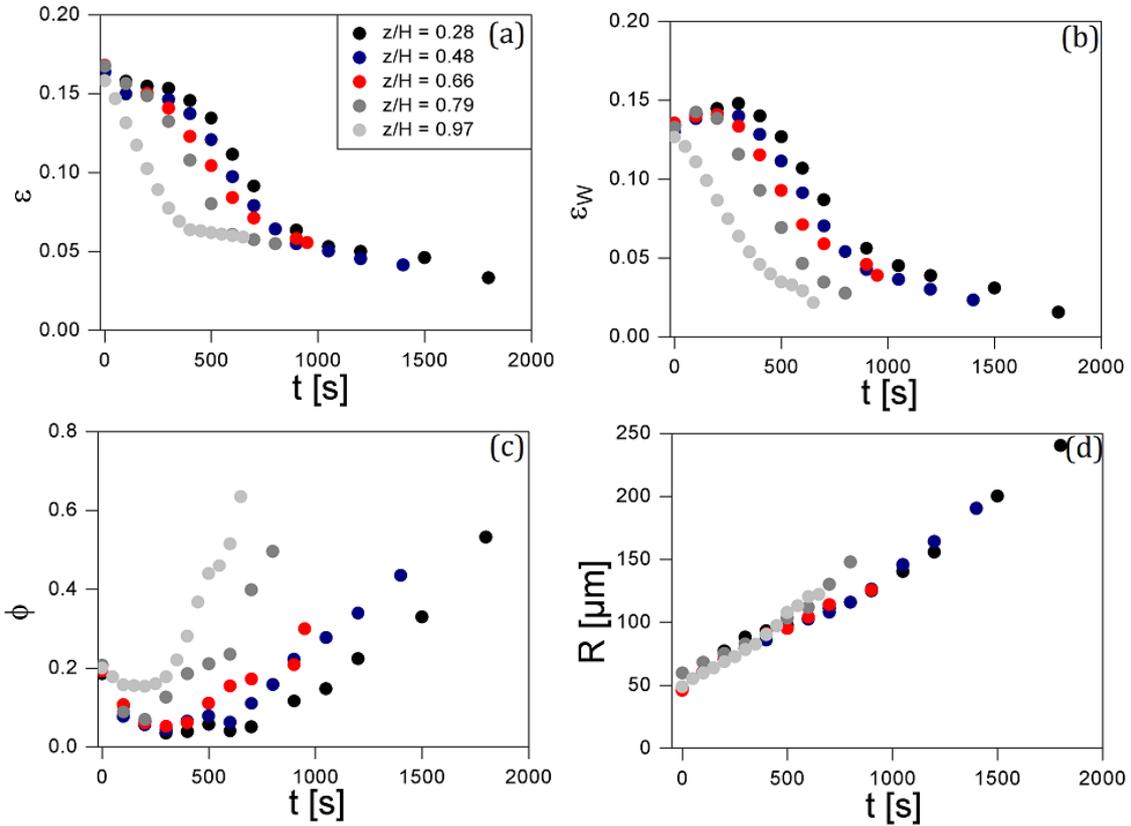

**Fig. 4** *Different fluid fractions as a function of time for foamulsions with an initial liquid fraction $\varepsilon_i$ = 0.15, different initial oil fractions $\phi$ (ranging from 0.1 to 0.4) and two different emulsions (SG and SN).* **(a)** *Total liquid fraction as a function of time.* **(b)** *Water fraction as a function of time.* **(c)** *Oil fraction in the emulsion as a function of time.* **(d)** *Time evolution of the average bubble radius. For all figures, open squares are for SN emulsions and circles are for SG emulsions. Black triangles are for pure SDS foams.*

Fig. 4(a) shows the total foam liquid fraction $\varepsilon$ as a function of time t measured using photographs of the surface of the column analyzed with *Eqn (3)*. The triangles show the case of an oil-free foam; $\varepsilon$ decreases continuously down to around 1% at t = 800 s. The measurements do not continue afterwards because the foam is completely destabilized. The time at which the foam disappears is defined as the life-time $t_f$. The other curves correspond to different oil fractions from 0.1 to 0.4, empty symbols are for SN foams ($r_H$ = 28 μm) and filled symbols are for SG foams ($r_H$ = 12 μm). The foams made with $\phi$ = 0.1 drain somewhat more slowly than the SDS foam, but as they break at a higher final liquid fraction, the foam life-time is very similar (around 800 s). As the initial oil fraction in the emulsion increases to 0.2 the drainage is slowed down further and e flattens off after around 800 s to a liquid fraction of about 0.08, which we will denote by $\varepsilon_B$. Although the initial drainage of SG and SN foams is very similar, their overall stability is different: the SG foam disappears shortly after plateauing at 1 000 s, while the SN foam persists for 3 000 s. The drainage curves for $\phi$ = 0.3 and 0.4 are qualitatively very similar to the $\phi$ = 0.2 curves. The foams drain until $\varepsilon$ reaches a plateau, at around 1 000 s and 2 000 s for foams made from emulsions with $\phi$ = 0.3 and 0.4 respectively. Again SN foams are more stable and have a life-time of almost 4 000 s compared with the 2 200 s of SG30 foams. A slight increase is measured in the liquid fraction of the SG40 foam after 2 500 s.

*Fig. 4(b)* shows the evolution of the water fraction inside the same foams, measured using conductivity. The initial water fractions are different due to the presence of different amounts of oil. The drainage is the fastest with pure SDS foam and slows down as $\phi$ increases. At the difference of $\varepsilon$, $\varepsilon_W$ does not level off as the foam ages.

*Fig. 4(c)* shows the oil fraction in the liquid phase as a function of time obtained with *Eqn (4)* from $\varepsilon$ and $\varepsilon_W$. The initial oil fractions used are in agreement with those in the emulsion which have been set during generation. In all samples the volume fraction of oil initially slightly decreases before starting to increase. The behaviour of SN and SG foamulsions at short times is very similar, but differences arise after 1 000 s when $\phi$ is about 0.3 or larger: the higher the initial oil fraction, the slower the increase of $\phi$ in time.

*Fig. 4(d)* shows the evolution of the average bubble size for the different foams. The initial bubble radius is around 30 μm. The initial growth is quite similar for all foams, and varies as $t^{0.5}$ (see *Fig. S3* (ESI†) for a log–log representation). This is as expected for dry foams, suggesting that coarsening is controlled by gas diffusion in the liquid phase as in oil-free foams [27].

### 4.4 Drainage at different heights

We have also studied the evolution of the drainage of the foamulsions at different heights with SG20 and $\varepsilon_i$ = 0.15. The measurements were taken at different heights in the column, z (z = 0 is the bottom of the column). The different measurements were taken at z/H = 0.28, 0.48, 0.66, 0.79 and 0.97, with H the total height of the foam column.

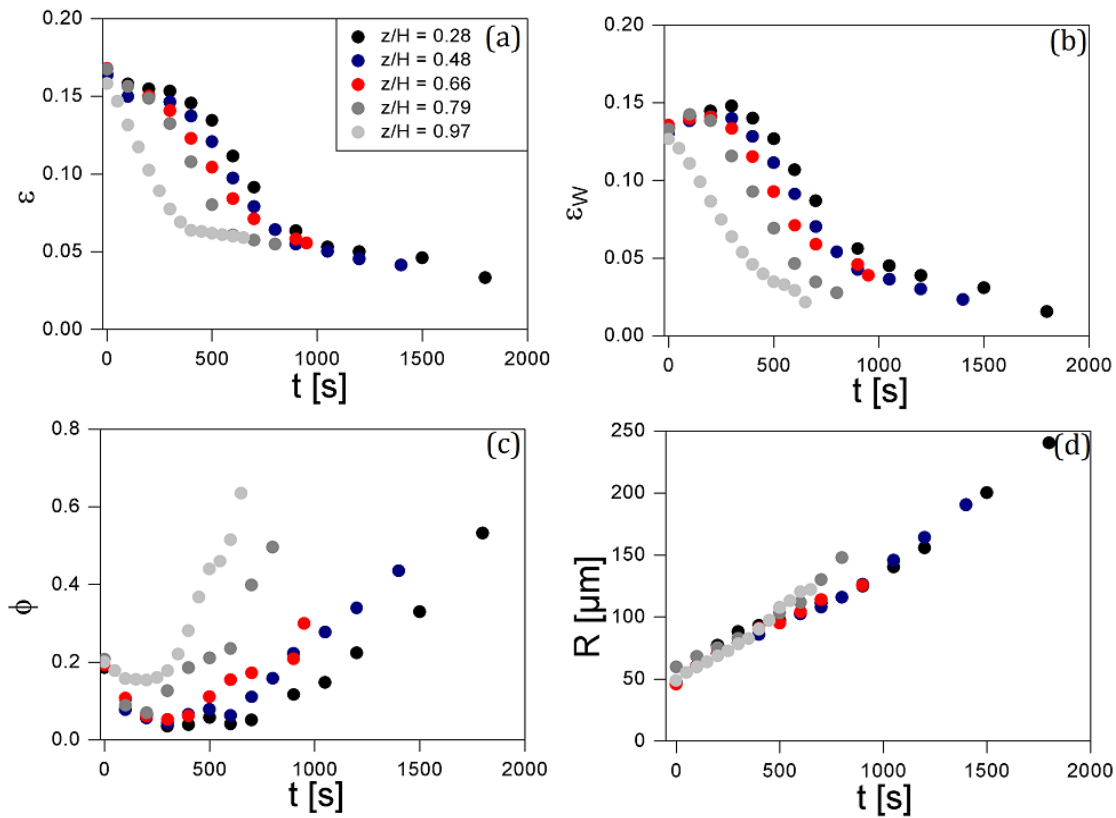

*Fig. 5* Drainage profiles at different heights for SG20 emulsions. **(a)** Liquid fraction as a function of time for different height ratios. Black point correspond to z/H = 0.28, blue points to z/H = 0.48, red points to

$z/H = 0.66$, gray points to $z/H = 0.79$ and light gray points to $z/H = 0.97$. *(b)* Water fraction as a function of time. *(c)* Oil fraction in the emulsion as a function of time. *(d)* Normalized average bubble radius as a function of time. A value of height ratio $z/H = 0$ corresponding to the bottom of the column and $z/H = 1$ being the top most position.

*Fig. 5(a)* shows the liquid fraction of the foamed emulsion as a function of time at the different heights in the foam. The light grey circles are taken very close to the top of the foam ($z/H = 0.97$), where the liquid fraction starts to decrease immediately after the generation of the foam. The decrease continues until around $\varepsilon = 0.06$ at which point drainage slows down considerably, e becoming almost constant. For $z/H = 0.79$, little drainage occurs in the first few hundred seconds after which the liquid starts draining at a rate similar to $z/H = 0.97$. For smaller $z/H$, the trends are similar with a lag time that increases when $z/H$ decreases in agreement with theoretical
predictions (Fig. 3). The behavior for $\varepsilon_W$ is similar to that of e, with a lag time increasing with decreasing z (*Fig. 5(b)*). The corresponding oil volume fractions in the liquid phase are shown in *Fig. 5(c)*. At the top of the foam, ɸ remains almost constant before starting to increase as drainage slows down (*Fig. 5(a)*). The oil concentration becomes very high and ɸ is above 60% when the foam destabilizes. In lower parts of the foam ɸ only slightly decreases before starting to increase. The increase occurs later and more slowly if z is smaller.

*Fig. 5(d)* shows the bubble radii at different positions in the foam. We see that the rate of coarsening is not significantly influenced by the liquid fraction. The same figure has been plotted in a log–log scale and is available in ESI.† Here again, the bubble growth follows the theoretical prediction as the data of *Fig. 4(d)*. However, the bubble growth is slightly faster at later stages near the top of the foam. This is maybe because of the onset of bubble coalescence.

## 5 Discussion

To explain what is happening in our foamed emulsions we propose a mechanism in three steps: a first step of fast drainage where both oil drops and air bubbles are moving wildly; a second step of emulsion drainage through the foam; and a third step during which drainage slows down considerably and the drops get blocked in the foam structure and can even cream. We describe the steps individually in the following.

### Step 1: fast homogeneous drainage: $t < t_0$

As soon as the cell is filled we start the measurements of the fluid fractions. We notice that lower down the foam there is a small delay time, defined as $t_0$ at the beginning, during which nothing happens. This delay time is longer the further down we are in the foam, as seen in *Fig. 5(a)*. We can shift the curves by t0 to scale them together, as shown in *Fig. 6(a)*, where we have plotted $\varepsilon$ as a function of $t / t_0$. Time t0 appears to measure the arrival of the initial drainage front, which advances with velocity V given by *Eqn (7)*. We have plotted t0 as a function of the height in the inset of *Fig. 6(a)*. Apart from the point at the top of the foam (z = 195 mm), the points are well aligned on a straight line indicating a constant velocity. We can calculate the velocity of the initial drainage front as $V_{exp} = z / t_0$, which gives 300 mm $s^{-1}$. The slope is the same both with (circles) and without oil (triangles), so with SG20 and with SDS.

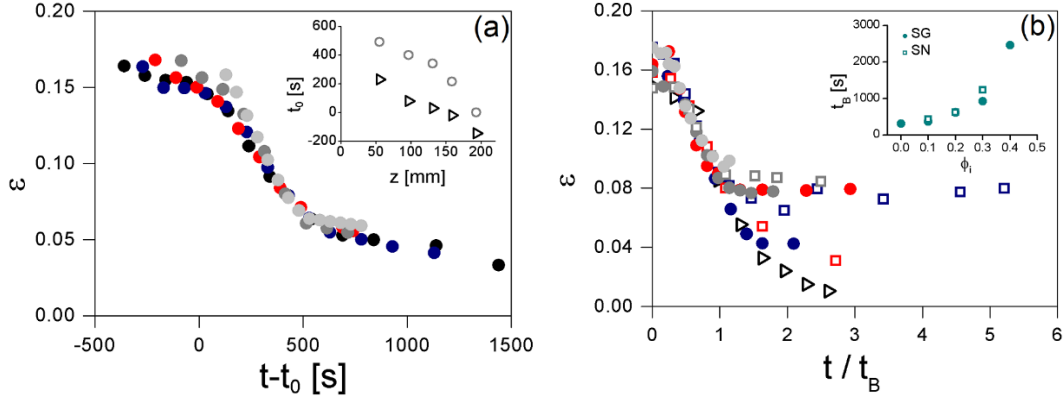

*Fig. 6 (a)* Liquid fraction as a function of time shifted by delay time $t_0$; inset: $t_0$ versus height for different oil fractions. Black triangles correspond to SDS foams and gray empty circles to SG20 foamed emulsions. *(b)* Liquid fraction as a function of time normalized by the scaling time t; inset: t as a function of initial oil fraction.

Let us first compare $V_{exp}$ with the prediction of *Eqn (7)*, where we use $K_d \sim 1/150$ [33], $L \sim 0.72\,R$, $R \sim 35\,\mu m$, $\eta = 5$ mPa s and $\varepsilon_i = 0.15$. We find $V_{exp} \sim 3\,\mu m\,s^{-1}$, two orders of magnitude slower than in our experiments. This is because the front velocity model is established for dry foams while ours are wet ($\varepsilon_i = 0.15$). This also means that we are probably not measuring a simple drainage front in the first instants, and indeed if we look at the foam during this time we see that it is not stationary. The bubbles are moving upwards and rearranging fast, while the model is for dry foams and has been experimentally verified with larger bubbles [33]. Therefore the first phase of drainage is also the time during which the foam is stabilizing, which takes 100 s of seconds and up to 1 000 s with the foam made from SG40. During this time the oil volume fraction remains constant, so the emulsion drains as a homogeneous fluid. This is because the emulsion creaming velocity (e.g. $V_{cream} \sim 3.7\,\mu m\,s^{-1}$ for SG20) is much slower than the drainage velocity ($V_{exp} \sim 300\,\mu m\,s^{-1}$ for SG20). The emulsion drops are simply swept along with the draining water.

### Step 2: slow / dependent viscous drainage: $t_0 < t < t_B$

During step 2 the total liquid fraction e decreases gradually in the foams. In *Fig. 6(a)* we see that once we take into account the lag time in the onset of drainage the second phase of drainage does not depend on height. However, it does depend on the oil volume fraction, as seen in *Fig. 4(a)*. The curves can be scaled together using a characteristic drainage time $t_B$, and the normalized drainage curves with different volume fractions of oil are shown in *Fig. 6(b)*. Until $t_B$, the data collapse onto a single curve, suggesting that a change in the volume fraction of oil simply changes the time scale of drainage, but not the underlying mechanism.

The different $t_B$ are plotted in the inset of *Fig. 6(b)* as a function of volume fraction of oil. The data with the two different emulsions group together, suggesting that the volume fraction of oil is indeed the control parameter. The viscosity of the emulsion phase is not sufficient to understand the slowing down of drainage as the viscosities of the two emulsions are different (see *Fig. S1*, ESI†), yet the drainage is very similar. We should of course keep in mind that the comparison is not so simple, given that the emulsions are shear thinning, and during the evolution of the foam the shear rate they experience is also changing. As drainage slows down, the emulsion could start to cream within the foam and thus slow down drainage.

During the first steps, the volume fraction of oil inside the foam remains almost constant, it decreases down to a shallow minimum before rising back to the initial value. After a period of drainage, it will suddenly slow down strongly or even arrest if the emulsion oil fraction is above 0.2, as we reach $t_B$.

### Step 3: arresting drainage $t_B < t < t_F$

In the foams made from higher initial oil volume fractions, the total liquid fraction will suddenly stop evolving, in *Fig. 6(b)* we can see that ε at which drainage arrests, defined as $\varepsilon_B$ depends on the initial oil fraction in the foam. The $\varepsilon_B$ are plotted in *Fig. 7*. The drainage can stop for different reasons, one is that the liquid fraction reaches its equilibrium value. This can be estimated for our foams as below 1% at a height of 17 cm and with a bubble radius of 100 μm using the expression from Ref. [29]. This is much smaller than what we observe and means that in our case the drainage does not stop because the equilibrium liquid fraction is reached. Although the total liquid fraction has reached a constant value, the water liquid fraction continues to decrease very slowly (see *Fig. 4(c)*), which means that the oil fraction in the continuous phase increases (see *Fig. 4(c)*). This means that the oil drops have started to block the channels, and they can continue to slowly cream. Solid particles have been shown to stop drainage by blocking in Plateau borders once the confinement parameter l becomes higher than one. At this point the size of the particles becomes larger than that of the Plateau borders, see *Section 3.1*. If we use the size of the individual drops (SG = 1.2 μm and SN = 0.3 μm) we find l that are between 0.03 and 0.4 throughout the drainage. These are much too small to cause blocking. However, we saw in Section 4.2 that the drops form clusters that are much larger. If we calculate l with the cluster sizes, we find values ranging from 1.8 to 4.4 immediately after the foams have been made. Therefore, the emulsion droplets should have been blocked from the start. However, this is not the case as seen in *Fig. 4(c) and 5(b)*.

In order to explain the sudden arrest of drainage we go back to the rheology of the emulsions, which are all shear thinning (see *Fig. S1*, ESI†). The viscosities at low shear rates are high for emulsions at the volume fractions studied, but this is because the emulsions are strongly flocculated. Therefore at small shear rates the loose clusters of droplets offer strong resistance to flow, but as the shear rate increases the clusters can be broken and the viscosity drops. In the foamed emulsions initial drainage is fast, and the emulsion clusters are deformed and broken. Their viscosity is low and drainage is homogeneous. However, as drainage slows down the clusters are much less sheared and start to block in the Plateau borders. The further slowing of drainage decreases the shear on the emulsions and drainage almost arrests. Flocculated silica particles have been studied using forced drainage, and similarly, at high flow rates they were shown to flow, but at lower imposed flow rates they jammed [32] Of course in the case of emulsions the clusters are blocked towards rising rather than falling due to the density difference. Once the liquid fraction becomes stable, the oil volume fraction continues to increase, probably because of creaming of the emulsion, and the foams eventually simply collapse. They collapse despite the high liquid fractions at which they can be blocked as seen in *Fig. 7*. Although SDS foams reach a final liquid fraction of around 0.01, foamulsions disappear at $\varepsilon_f$ between 0.04 and 0.1 depending on the ϕ. *Fig. 7* shows the values of $\varepsilon_f$. This suggests that at the end of the life of the foams, the emulsion turns from a stabilizing element to a destabilizing presence. The emulsion droplets act as antifoam particles.

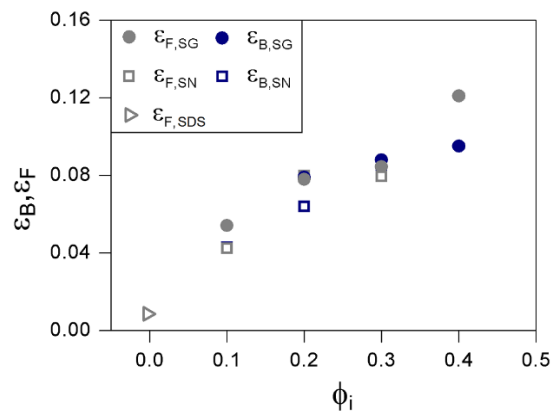

*Fig. 7* Blocking (gray points) and final (dark blue points) liquid fractions as a function of initial oil fraction in the emulsion. Circles refer to SG foamed emulsions, open squares to SN foamed emulsions and the open triangle to a pure SDS foam (oil free foam).

The equilibrium surface tensions and antifoam coefficients summarized in *Table 1* told us that rapeseed oil is a weak antifoam. However, the antifoam coefficients do not tell it all and the entry barrier is important [10]. There is a barrier to the drop – bubble coalescence which has to be overcome for the emulsion to act as antifoam. Slow antifoams often become efficient only once the foams have drained a little leading to a decrease in Plateau border size, and an increase in capillary pressure. Once the pressure is sufficiently high to overcome the barrier, the drops penetrate the interfaces and act as antifoam. In our foams, once the liquid fraction stops evolving, the Plateau borders will decrease in size as the bubbles grow. This leads to a higher capillary pressure, which combined with the increased concentration of oil droplets in the Plateau borders becomes sufficient to push the drops into the interfaces and the emulsion becomes antifoam.

The differences in the life-times of the SG and SN foams can be explained by the differences in entry barrier with drop size. Entry barrier is smaller for larger drops and the SG drops of 1.2 µm radius enter more easily than the SN droplets of 0.3 µm in radius. Therefore for the same initial oil volume fraction the foams made with smaller drops are much more stable.

## 6 Conclusion

We have studied the drainage of foamed emulsions. We have measured the concentrations of water, oil and air in a foam by combining measurements of electrical conductivity and surface photography. We show that the ageing process of foamulsions proceeds through a first phase of fast drainage, followed by a regime of slower drainage which eventually almost arrests, before a final slow collapse of the foam due to antifoaming action of oil.

During the first step, drainage is fast and both bubbles and drops rearrange continuously. Although foamed emulsions take longer to stabilise than oil-free foams, the stabilisation front velocity is the same for both. This drainage is so fast that emulsion drains homogeneously.

Once stabilised, in the second step liquid fraction decreases throughout the foam. This drainage is not only controlled by viscosity but also by initial oil volume fraction $\phi$, which remains constant. Drainage is fast enough to break clusters of oil drops that are initially formed in the emulsion. However, as drainage continues, the liquid fraction decreases leading to slower drainage. The clusters are less and less sheared and the dynamic viscosity of the continuous phase increases due to shear thinning properties of emulsions. Eventually the shear is so small that clusters can get trapped in the Plateau borders and the emulsions can cream.

In the next step, drainage is drastically slowed down, the liquid fraction e remains constant but oil fraction $\phi$ increases. This means that it is mainly water that drains, and the oil gets blocked and even creams. At some point, capillary pressure, that is the difference between liquid pressure and gas pressure in the bubbles, is high enough for the droplets to penetrate the interfaces and act as an anti-foaming agent: the foam collapses slowly. Foams made from smaller emulsion drops are much more stable, this is because the entry barrier for the droplets to break into the interface is higher for smaller drops. We have seen that the drainage of foamed emulsions proceeds in several steps, which depend on the size and concentration of the emulsion drops, as well as the interactions between them. This makes the evolution of the systems complicated and further work is required before detailed models or a full picture of the drainage process can be obtained. However the description of the drainage process that we offer is useful in understanding the destabilization of these complex, yet widely used systems.

# Glossary

| Variable | Name |
|---|---|
| $\varepsilon$ | Liquid fraction in the foam |
| $\varepsilon_i$ | Liquid fraction in the foam at initial time |
| $\varepsilon_W$ | Water fraction in the foam |
| $\varepsilon_A$ | Air fraction in the foam |
| $\varepsilon_O$ | Oil fraction in the foam |
| $\varepsilon_B$ | Blocking liquid fraction |
| $\varepsilon_F$ | Final liquid fraction |
| $\phi$ | Oil fraction in the continuous phase |
| $\phi_i$ | Emulsion volume fraction at initial time |
| $r$ | Average oil drop radius |
| $r_H$ | Hydrodynamic radius |
| $R$ | Average bubble radius |
| $R_{PB}$ | Average Plateau border radius |
| $\lambda$ | Confinement parameter |
| $\eta$ | Viscosity |
| $t_B$ | Blocking time, when drainage slows down |
| $t_F$ | Foam life-time |
| $t_0$ | Delay time in drainage due to height |
| $\gamma_{AW}$ | Air–water surface tension |
| $\gamma_{OW}$ | Oil–water surface tension |
| $\gamma_{AO}$ | Air–oil surface tension |
| E | Entering coefficient |
| B | Bridging coefficient |
| S | Spreading coefficient |

# Acknowledgements

We thank ESA (Soft Matter Dynamics) and CNES (Hydrodynamique des Mousses Humides) for financing this research.

# Supplementary information

## A. Emulsion viscosities

Emulsion viscosities η have been measured for different oil fractions ϕ and different shear rates $\dot{\gamma}$ using a Couette rheometer (PHYSICA MCR 300, Anton Paar) .The results are shown on the *Figure S1*.

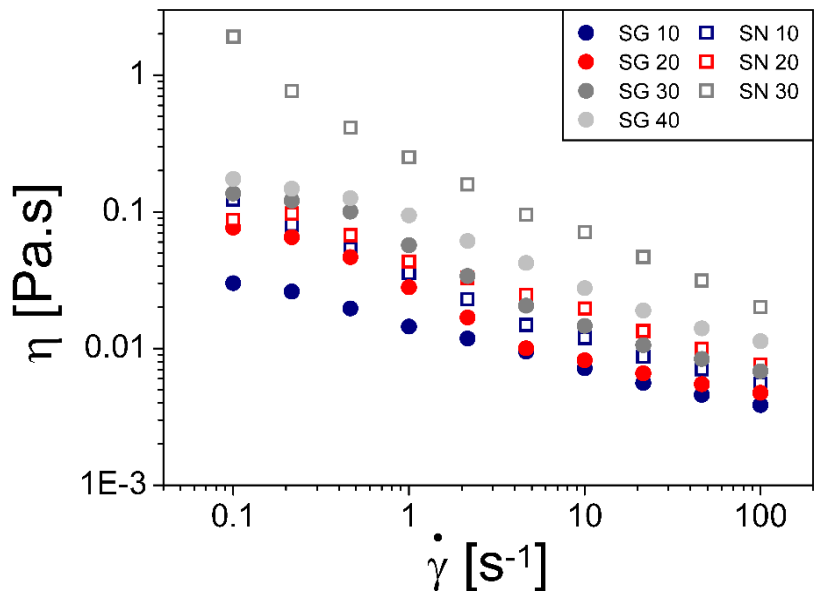

***Figure S1*** *Emulsion viscosity η as a function of shear rate $\dot{\gamma}$ for **(a)** Emulsions made using the double syringe technique (SG, colored circles ●). **(b)** Sonication (SN, colored open squares □). Points with the same color correspond to emulsions made with the same oil fraction. Names of samples refer to the generation technique followed by the oil volume fraction in the continuous phase.*

On Figure S1, we can see the emulsion viscosities as a function of shear rate $\dot{\gamma}$ for different emulsions. Viscosities decrease with $\dot{\gamma}$, the emulsions are shear-thinning. This means that during drainage as the drainage velocity and Plateau border size change leading to a change in the shear rate, the viscosity of the emulsion evolves. One can note that SN emulsions (small drops) are always more viscous than SG emulsions (bigger drops) and increasing the oil fraction increases the viscosity.

## B. Measuring oil fraction

In order to verify that our measuring method combining image analysis and conductivity is accurate, we have measured ϕ independently. We have centrifuged the foamulsions in order to remove the air and to separate oil and water. After centrifugation at 10 000 rpm for 10 minutes, the emulsion separates into water at the bottom and a concentrated emulsion at the top. Indeed, we are unable to induce coalescence of oil drops during centrifugation. We estimate the oil fraction between 0.64 (jammed drops) and 1 (coalesced drops). The oil fractions calculated from the ratio of water and concentrated emulsion volumes are shown in *Figure S2*. The red triangles are for fully coalesced emulsions ($\phi = 1$) and blue triangles for jammed spherical drops ($\phi = 0.64$).

These data from conductivity measurements are also shown in *Figure S2* (empty circles). They fall in between the calculated oil fractions for the extreme cases, meaning that the drops are partially distorted in the concentrated emulsion or that some coalescence has occurred, or both.

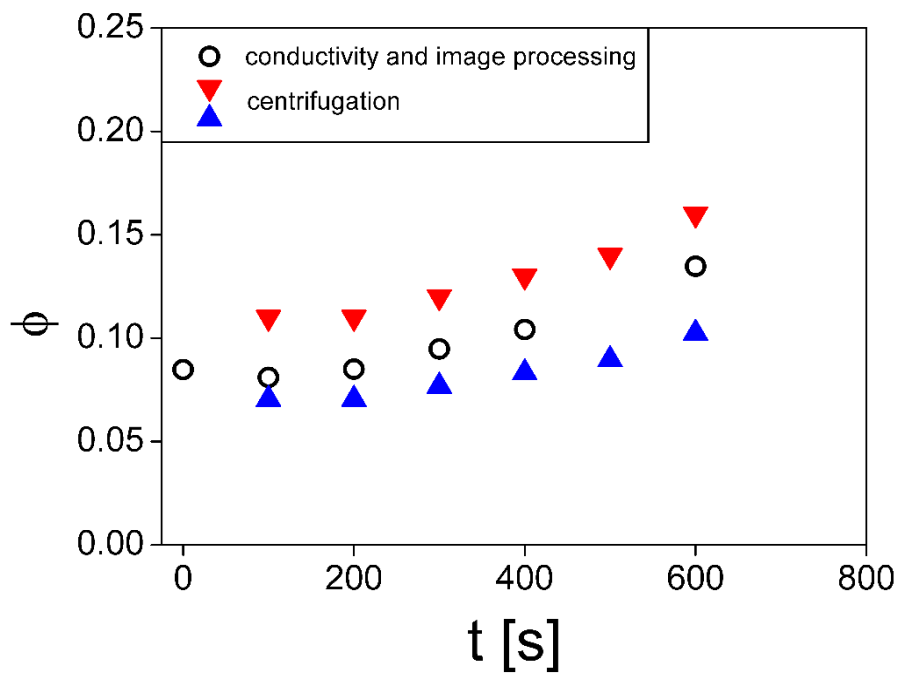

*Figure S2* Measured oil fraction in the emulsion versus foamulsion ageing time using electrical conductivity (black circles o) and centrifugation (red and blue triangles). Red triangles (▼) are the maximum values of the oil fraction and blue triangles (▲) are the minimum boundaries (jammed spherical oil drops).

These results suggest that our method is accurate and shows that as the foam ages, ϕ increases (see *Figures 4 and 5* of the main text).

## C. Coarsening

*Figure S3 (a)* presents the evolution of the average bubble radius R normalized by the radius at time $t = 0$. For the two sets of emulsions and for different initial oil fractions. *Figure S3 (b)* shows the results for a foamed emulsion containing 20 % of oil in the continuous phase and with an initial volume fraction of 0.15 at different heights. The data are plotted in log-log scale, to highlight that $R(t)$ evolves as $t^{1/2}$ as expected when coarsening is controlled by gas diffusion.

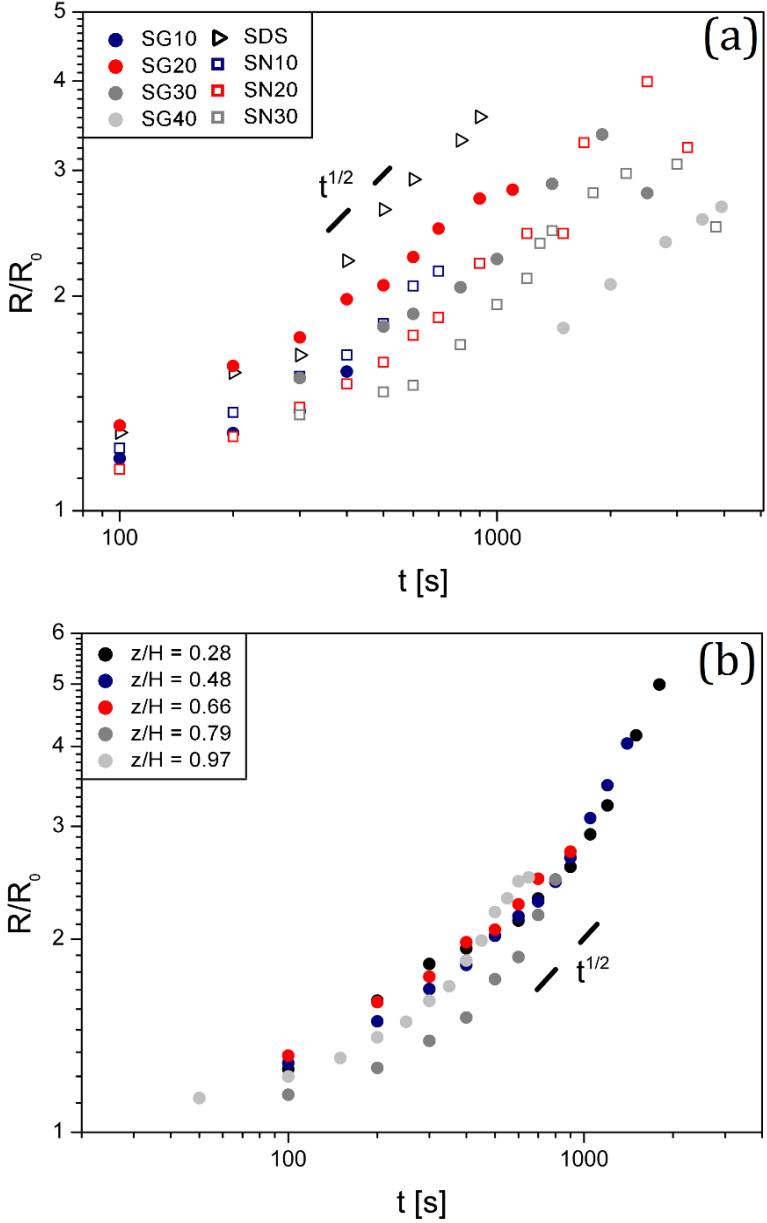

***Figure S3 (a)*** *Evolution in time of average bubble radius normalized by initial value for SN (□) and SG (●) foamed emulsions with initial oil fractions from 0 to 0.40.* ***(b)*** *Normalized average bubble radius as a function of time for a SG20 foamed emulsion at different heights in the column, 1 being the top most position in the column.*

## D. Confinement parameter λ

*Figure S4* presents the evolution over time of the confinement parameter λ, ratio of the hydrodynamic radius (cluster size) and Plateau border radius. In Figure S4 (a) the temporal evolution of λ is depicted for the different types of emulsions studied and in Figure S4 (b) at different heights in the foam for one emulsion (SG20 with 20% oil in the continuous phase). λ is greater than from the onset of measurements, yet the drops are blocked, suggesting that the clusters do not behave as hard spheres. They are likely somewhat distorted when they flow through the foam channels.

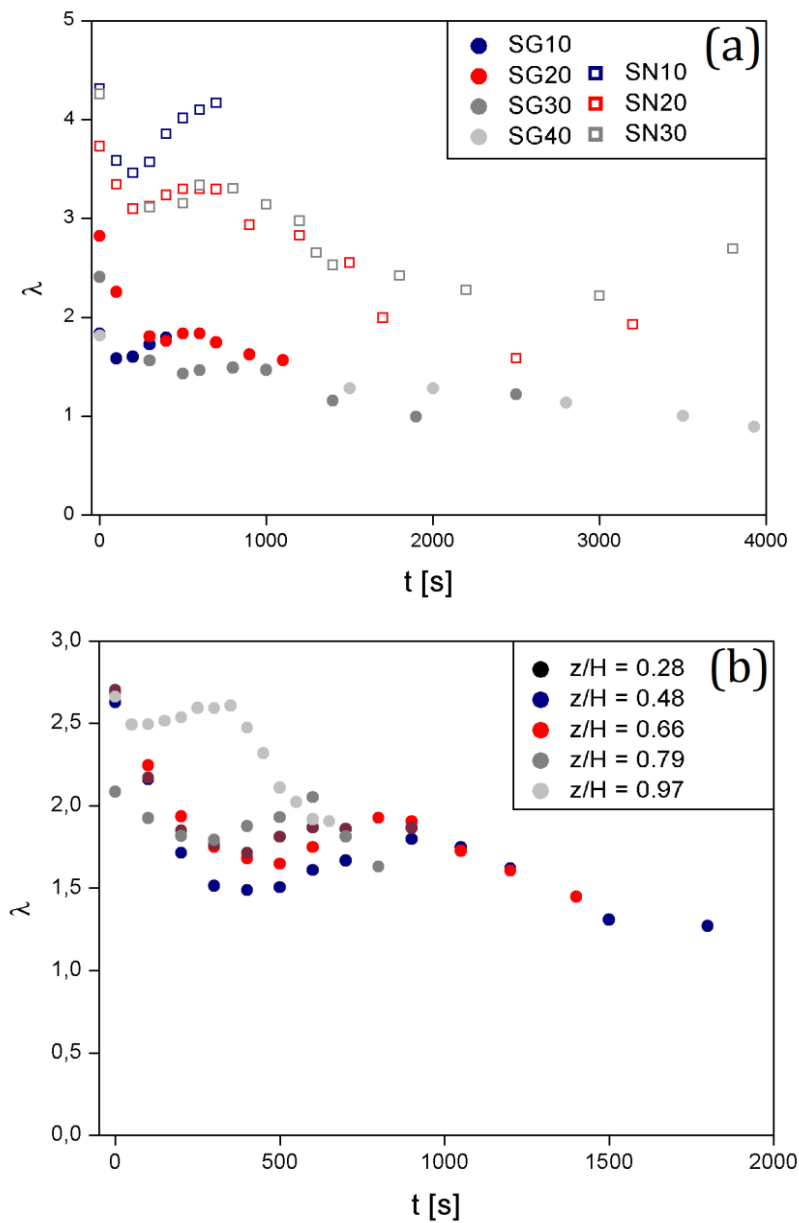

**Figure S4 (a)** *Evolution in time of confinement parameter λ for SN (□) and SG (●) foamed emulsions with initial oil fractions from 0 to 0.40.* **(b)** *Confinement parameter λ which is the ratio between oil drop aggregates and average Plateau border sizes, for SG20 foamed emulsion at different heights in the column as a function of time.*